\documentclass[12pt]{article}

\usepackage{scicite}
\usepackage{graphicx}
\usepackage{times}
\usepackage{color}

\newenvironment{sciabstract}{%
\begin{quote} \bf}
{\end{quote}}

\title{Observation of Poiseuille Flow of Phonons in Black Phosphorus}

\author
{Yo Machida,$^{1\ast}$ Alaska Subedi,$^{2,3}$ Kazuto Akiba,$^{4}$ Atsushi Miyake,$^{4}$\\ Masashi Tokunaga,$^{4}$
Yuichi Akahama,$^{5}$ Koichi Izawa,$^{1}$ and Kamran Behnia$^{6,7\ast}$\\
\\
\normalsize{$^{1}$Department of Physics, Tokyo Institute of Technology, Meguro 152-8551, Japan}\\
\normalsize{$^{2}$Centre de Physique Th\'eorique, \'Ecole Polytechnique, CNRS, Universit\'e Paris-Saclay,}\\
\normalsize{F-91128 Palaiseau, France}\\
\normalsize{$^{3}$Coll\`ege de France, 11 place Marcelin Berthelot, 75005 Paris, France}\\
\normalsize{$^{4}$The Institute for Solid State Physics, University of Tokyo, Kashiwa 277-8581, Japan}\\
\normalsize{$^{5}$Graduate School of Material Science, University of Hyogo, Kamigori 678-1297, Japan}\\
\normalsize{$^{6}$Laboratoire Physique et Etude de Mat\'{e}riaux (CNRS-UPMC), ESPCI Paris,}\\
\normalsize{PSL Research University, 75005 Paris, France}\\
\normalsize{$^{7}$II. Physikalisches Institut, Universit\"{a}t zu K\"{o}ln,  50937 K\"{o}ln, Germany}\\
\\
\normalsize{$^\ast$To whom correspondence should be addressed; E-mail:  machida@ap.titech.ac.jp;}\\
\normalsize{kamran.behnia@espci.fr.}
}

\date{July 2, 2018}

\begin{document}


\baselineskip24pt


\maketitle

\noindent
{\bf One-sentence summary:}\\
Hydrodynamic flow of phonons, previously detected only in a handful of solids, is observed in black phosphorus.

\clearpage


\begin{sciabstract}
The travel of heat in insulators is commonly pictured as a flow of phonons scattered along their individual trajectory.
In rare circumstances, momentum-conserving collision events dominate, and thermal transport becomes
hydrodynamic. One of these cases, dubbed the Poiseuille flow of phonons, can occur in a temperature window
just below the peak temperature of thermal conductivity. We report on a study of heat flow in bulk black phosphorus
between 0.1 and 80 K. We find a thermal conductivity showing a faster than cubic temperature dependence
between 5 and 12 K. Consequently, the effective phonon mean free path shows a nonmonotonic temperature
dependence at the onset of the ballistic regime, with a size-dependent Knudsen minimum. These are hallmarks of
Poiseuille flow previously observed in a handful of solids. Comparing the phonon dispersion in black phosphorus
and silicon, we showthat the phase space for normal scattering events in black phosphorus is much larger. Our results
imply that the most important requirement for the emergence of Poiseuille flowis the facility ofmomentum exchange
between acoustic phonon branches. Proximity to a structural transition can be beneficial for the emergence of this
behavior in clean systems, even when they do not exceed silicon in purity.
\end{sciabstract}

\section*{INTRODUCTION}
The finite thermal resistivity of an insulating solid is a manifestation of the anharmonicity of the underlying lattice. The most common approach to calculate the thermal conductivity of a given solid is to assume the harmonic approximation and introduce a finite lifetime for phonons that captures the effects beyond the harmonic approximation. In the Boltzmann-Peierls picture, heat-carrying phonons are scattered by other phonons or by crystal imperfections and boundaries. The magnitude of thermal resistivity is set by the rate of collisions relaxing the momentum of the traveling phonon. Only a subset of phonon-phonon scattering events, called Umklapp (or U) process degrade the heat current. The initial and final momenta in a U process differ by a multiple of a reciprocal lattice vector. On the other hand, a normal (or N) phonon-phonon collision cannot lead to thermal resistance in absence of Umklapp scattering. In an infinite sample, if all collisions were normal, the flow of phonons is expected to be undamped in the absence of any external field applied to sustain it~\cite{gurzhi}.

Black phosphorus (BP) has attracted much recent attention as a cleavable solid with a promising exfoliating potential towards two-dimensional phosphorene~\cite{xia,ling}. It has an orthorhombic crystal structure with puckered honeycomb layers in the $ac$ planes and van der Waals coupling between the layers (Fig.~1A and 1B). Unlike graphene, it has a tunable direct band gap ranging from 0.3 eV in bulk  to 2 eV in monolayer~\cite{das}. These features make BP a promising material for applications. In addition, the presence of a significant in-plane anisotropy  may  induce spatially-anisotropic transport~\cite{fei} absent in other graphene-like materials. Although electronic conduction in BP has been extensively investigated~\cite{akahama,akiba,xiang,akibaHall}, few studies have been devoted to its thermal transport~\cite{slack,flores,wang,sun,smith,hu}. They are restricted to relatively high temperatures and did not explore the low--temperature range, the focus of the present study.

Here, we report on a study of  thermal conductivity  along the $a$ and $c$ axes of BP single crystals and establish the magnitude and temperature dependence of the thermal conductivity down to 0.1 K. We find a moderately anisotropic thermal conductivity mainly reflecting the anisotropy of the sound  velocity. Unexpectedly, we find that below its peak temperature, the thermal conductivity evolves faster than the ballistic $T^3$. This feature, combined with size-dependence of the Knudsen minimum in the effective phonon mean-free-path, provides compelling evidence for Poiseuille phonon flow, a feature previously restricted to a handful of solids~\cite{beck}. We argue that this arises because of the peculiar phonon dispersions enhancing the phase space for normal scattering events in BP. This observation has important consequences for the ongoing research in finding the signatures of hydrodynamic phonon flow in crystals.

\section*{RESULTS}
Our samples are insulators, as seen in Figure~1C, which displays the temperature dependence of the electrical resistivity $\rho$ along the $a$ and $c$ axes. For both directions, $\rho$  shows a peak around 250 K, reflecting the change in the number of thermally-excited carriers across more than one gap. The magnitude of the resistivity, its temperature dependence, and its anisotropy are in agreement with an early study~\cite{akahama}. Below 50 K, $\rho$ displays an activated behavior, as seen in the upper inset of Fig.~1C. The activation energy of hole-like carriers is set by the distance in the energy between the top of the valence band and the acceptor levels~\cite{akahama}. Using Arrhenius equation $\rho=\exp(E_g/2k_{B}T)$ between 40 K and 20 K, one finds an activation energy of $E_g\sim$ 15 meV, in agreement with the previous result~\cite{akahama}. Below 20 K, the temperature dependence of $\rho$ becomes appreciably weaker, indicating that the system enters the variable range hopping (VRH) regime  where the electrical transport is governed by the carriers trapped in local defects hopping from one trap to another. One can describe the resistivity  by the expression {$\rho\propto\exp[(T/T_0)^{-1/(d+1)}]$ in a system with dimension $d$. As seen in Figs.~1D to 1F, it is hard to distinguish
the hopping space dimensionality from our results. The resistivity anisotropy  ratio, $\rho_{a}/\rho_c$ (lower inset of Fig.~1C), shows little temperature dependence in the activated regime. The ratio, almost constant ($\sim$ 3.5) in the activated regime and  reflecting the intrinsic anisotropy in the carrier mobility~\cite{akahama}, sharply increases in the VRH regime and attains $\sim$ 10.

Because of the extremely low electrical conductivity, heat is almost entirely transported by phonons, and we are going to focus on them. Figure~2A shows the temperature dependence of the thermal conductivity, $\kappa$, along the $a$ and $c$ axes in a BP single crystal. Thermal conductivity of bulk BP was recently measured by four different groups ~\cite{wang,sun,smith,hu}. Sun {\it et al.}~\cite{sun} found a large and anisotropic thermal conductivity, much larger than what was found in an early polycrystal study by Slack~\cite{slack} and a single crystal study by another groups~\cite{wang,hu}. As one can see in the figure, our data for the samples of $\sim$ 100 $\mu$m thickness (solid circles) is in good agreement with the data reported by Sun {\it et al.}~\cite{sun}, including the anisotropy essentially due to the sound velocity (see the Supplementary Materials for details).  Both sets of data point to an intrinsic bulk conductivity much larger than what was found in a polycrystal~\cite{slack}. The large thickness dependence of the thermal conductivity in BP found by our study (see below), in agreement with what was recently reported by Smith {\it et al.}~\cite{smith}, provides a clue to the striking discrepancy between the single-crystal and polycrystal data. If, even at temperatures exceeding 100 K, the phonon thermal conductivity is affected by sample dimensions, then the attenuation of thermal conductivity in the presence of grain boundaries is not surprising.

In absence of anisotropy, the lattice thermal conductivity can be represented by the following simple expression:
\begin{equation}
\kappa=\frac{1}{3}C\langle v\rangle l_{ph},
\end{equation}
where $C$ is the phonon specific heat, $\langle v\rangle$ is the mean phonon velocity, and $l_{ph}$ is the phonon mean-free-path. Note that the 1/3 prefactor, a consequence of averaging over the whole solid angle in isotropic medium, can be different in presence of anisotropy. At temperatures exceeding the Debye temperature, where the Dulong-Petit law applies, the specific heat saturates to a constant value. In this regime, the temperature dependence of the thermal conductivity is governed by the temperature dependence of the phonon mean-free-path, which becomes shorter with increasing temperature.  At the other extreme, i.e. at low temperature, the mean-free-path saturates to its maximum magnitude, of the order of the sample dimensions, and the temperature dependence of thermal conductivity is set by the $T^3$ behavior of the phonon specific heat. Between these two extreme regimes, dubbed the kinetic (at high temperature) and the ballistic (at low temperature), the thermal conductivity passes  through a peak. As demonstrated by  extensive studies on  LiF~\cite{thacher}, introducing isotopic impurities  damps  $\kappa_{max}$, the peak  value of the thermal conductivity. On the other hand, decreasing the size of the sample shifts the position of $\kappa_{max}$ with little effect on the magnitude of thermal conductivity in temperatures exceeding the peak.

Our key finding is the observation of a $\kappa$ evolving faster than $T^3$ just below the peak temperature in  BP. This can be seen in Fig.~2B.
As a consequence, $\kappa$ divided by $T^3$ shows a maximum or a shoulder around 10 K (see the inset of Fig.~2A).
On further cooling, the heat transport eventually becomes ballistic as evidenced by the $T^3$ behavior of $\kappa$ (saturation of $\kappa$/$T^3$) observed in the low-temperature data of the thickest sample (open circles in Fig.~2A). As a consequence, the effective phonon mean-free-path, $l_{ph}$, extracted from Eq.~(1), presents a local maximum (around 10 K) and a local minimum (around 4 K) (solid circles in Fig.~3C and 3D). Here,  $l_{ph}$ was calculated using  $\langle v_a\rangle$ = 0.536 $\times$ 10$^4$ m/s, $\langle v_c\rangle$ =  0.354 $\times$ 10$^4$ m/s (these are calculated velocities in agreement with previous calcalculations~\cite{kaneta} and neutron scattering measurements~\cite{fujii}) and the measured specific heat of the crystal (see the Supplementary Materials for details). 
In millikelvin temperatures, $l_{ph}$ becomes comparable to the sample thickness (see the inset of Fig.~3C).
Decades ago, the phonon Poiseuille flow~\cite{beck} was detected in Bi~\cite{kopylov} by observing these two features, namely a faster than $T^3$ evolution of $\kappa$ (Fig.~2C) and a peak in the temperature dependence of $l_{ph}$ just below the maximum thermal conductivity (Fig.~3E). Similar features have also been found in crystalline $^{4}$He ~\cite{deglin}, $^{3}$He~\cite{thomlinson}, and H~\cite{zholonko}.

In order to exclude any experimental artifact, we measured the thermal conductivity of P-doped Si using the same experimental setup. As seen from Fig.~2A, the temperature dependence of $\kappa$ for P-doped Si  conforms to what was found previously~\cite{glass}. $\kappa$ displays a cubic temperature dependence between 2 K and 5 K. (Fig.~2D). 
To show this point more explicitly, $l_{ph}$ calculated from Eq.~(1) is plotted as a function of temperature in the inset of Fig.~3D. The physical parameters used in the calculation are $\theta_{D}$ = 674 K and $\langle v\rangle$ = 6700 m/s~\cite{slackSi}. As expected, $l_{ph}$ does not show a local maximum, and at low temperature, its magnitude is in reasonable agreement with the sample dimensions (0.3 $\times$ 1.4 $\times$ 4.0 mm$^3$).

Only Umklapp and impurity scattering events degrade the momentum of the traveling phonons. Normal scattering events conserve crystal momentum and do not contribute to the thermal resistance. When normal scattering is sufficiently strong and the momentum-degrading scattering  events are significantly rare, normal scattering can even enhance the heat flow~\cite{gurzhi,beck}. The essential condition for this is given by  the inequality:
\begin{equation}
l^N\ll d\ll l^R.
\end{equation}
Here $d$ is a characteristic sample dimension, and $l^N$ and $l^R$ are normal and resistive mean-free-paths, respectively. In such a situation, phonons lose their momentum only by diffuse boundary scattering,  and they flow freely  while continuously exchanging momentum in the center of substances, analogous to the Poiseuille flow in fluids (Fig.~2E). As a result, moving like a Brownian particle, a phonon traverses the path length of the order of $l_{ph}\sim d^2/l^N$ between successive collisions with the boundaries, and in the ideal case $l_{ph}$ can reach a value larger than the characteristic sample dimension $d$. If $l^N$ increases with decreasing temperature as $T^{-5}$, then since $\kappa\sim C\langle v\rangle d^2/l^N$, thermal conductivity is expected to follow as $T^8$~\cite{gurzhi}. 
Experimentally, in all systems in which Poiseuille flow has been reported (namely Bi, solid He, and H), what has been observed is a faster than $T^3$ thermal evolution of $\kappa$ and a non-monotonic $l_{ph}(T)$ ~\cite{deglin,thomlinson,zholonko,kopylov}, not a superlinear size dependence. This is also the case in the present study on BP.

In the narrow temperature window of the Poiseuille flow, the dominance of Normal scattering creates an unusual situation where warming the system enhances the  mean-free-path, because it enforces momentum exchange among phonons. In other words, the temperature dependence of thermal conductivity is set by the temperature dependence of the phonon viscosity and not exclusively by the change in the population of thermally-excited phonons or the frequency of momentum-relaxing events, as it is the case in the ballistic and diffusive regimes.  At sufficiently low temperatures, the phonons turn to behave just like a Knudsen flow of gas~\cite{deglinBi}. At this transition from the Poiseuille flow to the Knudsen flow, $l_{ph}$ is accompanied by a mean-free-path minimum, dubbed Knudsen minimum, which occurs when  $d/l^{N}\sim$ 1. This is where the diffuse boundary scattering rate is effectively increased due to normal scattering (Fig.~4A). 

By changing the thickness of the samples along the $b$ axis, we also examined a size-dependence of the thermal conductivity and confirmed that the effect is more prominent in larger samples (Fig.~3A and 3B). As expected, in thicker BP samples, the maximum $l_{ph}$ is longer. Moreover, as seen in the case of Bi (Fig.~3E)~\cite{kopylov}, both the local maximum and the local minimum in $l_{ph} (T)$ persist (Fig.~3C and 3D). Note also that, despite the anisotropic thermal conductivity, $l_{ph}$ attains values comparable to the sample dimensions at low temperatures irrespective of the orientation of the heat current, meaning that $l_{ph}$ is quasi-isotropic along the three principle axes and set by the average sample size.
This is derived from the low anisotropy in the sound velocity between the $ac$-plane and out-of-plane orientations in spite of the layered structure of BP (see Fig.~S4 and Table~S2 in supplementary materials). Remarkably, the peak thermal conductivity in larger BP crystals attains a magnitude as large as 800 W/Km, much higher than the values found in thinner samples~\cite{smith} (Fig.~3A and 3B), but still smaller than the maximum thermal conductivity of large ($\sim$ cm) crystals of Si (3000 W/Km)~\cite{glass} or Bi (4000 W/Km)~\cite{kopylov}.

In the original Gurzhi picture~\cite{gurzhi}, the quadratic dependence of $l_{ph}$ on the sample dimension, $d$, would lead to a superlinear size-dependence of the thermal conductivity in the Poiseuille regime. As one can see in Fig.~5B, this is absent in BP, as it was the case of Bi~\cite{kopylov}. Now the Poiseuille flow of phonons is expected to emerge concomitantly with the second sound, a wave-like propagation of entropy~\cite{beck,rogers}. Both require a peculiar hierarchy of scattering events~\cite{beck,cepellotti}, where normal scattering rate is much larger than boundary scattering and the latter much larger than resistive (Umklapp and impurity) scattering. Its experimental observation in Bi in the same temperature range~\cite{narayanamurti} confirms the interpretation of static thermal conductivity data.

The absence of superlinear size dependence can be traced back to the expression for phonon kinematic viscosity, proposed by Gurzhi; $\nu$ = $v_Tl^N$ ~\cite{gurzhi}, which implies that the phonon fluid is non-Newtonian. Here, $v_T$ is the phonon thermal velocity whose magnitude is comparable to the velocity of the second sound, so that $v_T\sim$ $\langle v\rangle$/$\sqrt{3}$ = 2000-3000 m/s and $l^N$ is estimated to be of the order of 10~$\mu$m at the Knudsen minimum. This yields $\nu \sim$ 0.02-0.03 m$^2$/s, several orders of magnitude larger than the kinematic viscosity of water. This comparison is to be contrasted with the case of electrons in PdCoO$_{2}$~\cite{moll} whose dynamic viscosity was found to be comparable to the dynamic viscosity of water. Since the phonon viscosity  depends on the normal scattering rate and local velocity, it is not a constant number at a given temperature. The velocity profile of such a non- Newtonian fluid is much flatter than parabolic~\cite{eu}. This makes the size dependence much less trivial than in the parabolic and Newtonian case. A recent theoretical study~\cite{ding} has shown that the thickness dependence can be either sublinear or a superlinear according to the relative weight of boundary and normal scattering rates.

There is however another signature of the Poiseuille flow in the thickness dependence, which has been detected by our experiment.  In the Poiseuille regime, boundary scattering is rare. Therefore, with increasing thickness, phonons can travel a longer distance between successive collisions with the boundaries, giving rise to an enhancement of the thermal conductivity and the maximum in $l_{ph}$ with the thickness. On the other hand, at the onset of the ballistic regime, the fraction of phonons which contribute to the diffuse boundary scattering due to numerous normal scattering increases with the thickness. As a consequence, the Knudsen minimum shifts to lower temperatures in the thicker samples (Fig.~4A). One can  see both these features in a plot of $l_{ph}$ normalized by its minimum value $l_{ph}^{min}$, clearly along the $c$ axis (Fig.~4C), but not clearly along the $a$ axis (Fig.~4B) as it was the case of Bi~\cite{kopylov} (Fig.~4D). We conclude that the Poiseuille flow is most prominent along the $c$ axis. This suggests that the relative weights of Normal and Resistive scattering rates depend on orientation and one needs to go beyond a Boltzmann picture with a scalar scattering time~\cite{broido}.

The Ziman regime is another hydrodynamic regime~\cite{beck}. Here normal scattering still dominates, but resistive scattering becomes larger than boundary scattering. In this regime, expected to occur just above the peak temperature, one expects $\kappa$ to decrease exponentially with increasing temperature before displaying a  $T^{-1}$ temperature dependence at higher temperatures. Our measurements do not detect a regime where the thermal conductivity follows an exponential behavior in BP. Instead, all samples displayed a robust $T^{-1}$ behavior in a wide temperature window extending down to 40 K, an order of magnitude lower than the Debye temperature (Fig.~5A). Moreover, the slope of the $T^{-1}$ temperature dependence was larger in the thicker samples (See the inset of Fig.~5A).

Changing the thickness from 6(15) to 20(30) microns of the $a$($c$) -axis sample is to multiply the number of parallel heat-conducting phosphorene layers by a factor of three(two). Surprisingly, while no change in phonon dispersion is expected to occur in such a context, the enhancement in thermal conductivity persists up to 80 K, the highest temperature of our range of study (Fig.~5B).  Such a large thickness dependence (first reported in sub-micronic samples by Smith {\it et al.}~\cite{smith}) is unusual, as shown by a comparison with the much more modest effect observed in Ge~\cite{ziman}.  In Bi, a comparable enhancement is seen in its Poiseuille regime (Fig.~5B), but it rapidly dies away with heating. A plausible explanation is that in BP some heat carriers  remain  ballistic at high temperature. Interestingly, previous theoretical treatments of the thermal conductivity in graphene and graphite have argued for the presence of normal processes up to room temperature leading to collective phononic modes with a mean-free-path of the order of hundreds of micrometers \cite{leeHydro,cepellotti,fugallo}. Our observation may also indicate the presence of substantial normal processes above the peak temperature of thermal conductivity in BP.

\section*{DISCUSSION}
Our theoretical calculations confirm that phonon-phonon scattering in BP is large at low energies. The phonon dispersions of BP and Si are shown in Fig.~6, which agree well with previous calculations~\cite{gian91,zhan16}.  The phonon dispersions of Si show highly dispersive acoustic branches that lead to high thermal conductivity. The acoustic branches of  BP are less dispersive due to a relatively weaker bonding than the one in Si, which has strong $sp^3$ bonds. This leads to a higher phonon density of states (PDOS) in the low-frequency region,  which allows for a larger phase space for the momentum-conserving normal three-phonon scattering events in BP.

The calculated PDOS of BP and Si are shown in Fig.~6G and 6H. The PDOS of BP is significantly larger than that of Si in the low-frequency region below 100 cm$^{-1}$ as a consequence of the smaller dispersion of the acoustic branches. The phonon scattering processes involving the linearly dispersive acoustic modes conserve momentum and are not resistive to heat flow. The large low-frequency PDOS in BP provides much more phase space for these non-dissipative scattering events. This can favor the emergence of Poiseuille flow by establishing the required hierarchy of time scales as discussed above.

In summary, by detecting that phonons present a faster than $T^{3}$ thermal conductivity in a finite temperature window, we have identified a hallmark of Poiseuille flow of phonons in BP, which was previously seen only in four other solids~\cite{footnote}. Theoretical calculations indicate that the peculiarities of phonon dispersion leads to a significant enhancement of the phonon-phonon normal scattering in BP compared to an archetypal insulator such as Si. We note that BP is close to a structural instability. Like Bi and other column V elements, its crystal structure results of a slight distortion of the cubic structure~\cite{littlewood,behnia} Our results indicate that hydrodynamic phonon flow can be observed in a system close to a structural instability~\cite{footnote}.

\section*{MATERIALS AND METHODS}
Single crystals of BP were synthesized under high pressure~\cite{endo}. 
Magneto-phonon resonance, which requires reasonable purity, was observed in a sample from the batch used in the present experiments~\cite{akiba}.
A single crystal of P-doped Si was provided by Institute of Electronic Materials Technology (Warsaw).
The electrical resistivity and the thermal conductivity measurements were performed along the $a$ and $c$ axes of BP. Each sample has a rectangular shape with edges parallel to the three high-symmetry axes; the $a$, $b$, and $c$ axes. Length of the samples are summarized in Table S1. The $a$($c$) axis sample has the largest length along the $a$($c$) axis, respectively. 
The thickness dependence of the thermal conductivity was investigated by using the 
same sample.
The thickness (number of layers) along the $b$ axis was decreased during the course of the investigations by cleaving.

The thermal conductivity was measured by using a home built system. We employed a standard one-heater-two-thermometers steady-state method. 
A very similar setup was used to measure the thermal conductivity of 
cuprate~\cite{nakamae} and heavy-Fermion samples of
a comparable size~\cite{UPt3,YRS}.
The measurements were performed in the temperature range between 0.1 K and 80 K.
The thermometers, the heater, and the cold finger were connected to the sample by gold wires of 25 $\mu$m diameter. The
gold wires were attached on the BP sample by Dupont 4922N silver paste and were soldered by indium on the P-doped Si sample. The contact resistances were less than 10 $\Omega$ at room temperature. The temperature difference generated in the sample by the heater was determined by measuring the local temperature with two thermometers (Cernox resistors in the $^4$He cryostat and RuO$_2$ resistors in the dilution refrigerator). The heat loss along manganin wires connected to the two thermometers and the heater are many orders of magnitude lower than the thermal current along the sample including the thinner ones (Fig.~S1)}. The heat loss by radiation is negligible in the temperature range of our study ($T<$ 80~K), since it follows a $T^4$ behavior. 
The heat loss by residual gas is also negligible because the measurements were carried out in an evacuated 
chamber with a vacuum level better than $10^{-4}$ Pa. Moreover, the external surface of the chamber is directly in the helium bath at 4.2~K for the measurement above 2~K, so
that residual gas is condensed on the cold wall.
The experimental uncertainty in the thermal conductivity arising from the measurements of the thermometer resistances
is less than 1 $\%$ in the whole temperature range.
The main source of uncertainty results from uncertainty in the measured thickness of the samples, which is about 5 $\%$ in the 
thinnest sample.
The increase of thermal conductivity with the sample thickness, however, dominates over the experimental uncertainty.

The same contacts and wires were used for the electrical resistivity measurements by a four-point technique. A DC electric current was applied along the sample using the manganin wire attached on the heater. The Keithley 2182A nanovoltmeter was employed for the measurement of electrical voltage. The input impedance of the nanovoltmeter is larger than 10 G$\Omega$, which is well above the resistance of the sample even at the lowest temperatures ($R\sim$ 5~M$\Omega$).

\section*{SUPPLEMENTARY MATERIALS}
Supplementary material for this article is available at http://advances.sciencemag.org/cgi/
content/full/4/6/eaat3374/DC1\\
Supplementary Text\\
fig. S1. Thermal resistance of black phosphorus samples and manganin.\\
fig. S2. Reproducibility of thermal conductivity data.\\
fig. S3. Specific heat.\\
fig. S4. Phonon dispersions at low energies.\\
table S1. Size of samples.\\
table S2. Phonon velocity along the high-symmetry axes.\\
References \textit{(46-55)}

\noindent

\textbf{Acknowledgments: }
We acknowledge R. Nomura for fruitful discussions. 
\textbf{Funding: }This work was supported by JSPS Grant-in-Aids KAKENHI 16K05435, 15H05884, and 17H02920, Fonds ESPCI, and the European Research Council grant ERC-319286 QMAC. The calculations were performed at the Swiss National Supercomputing Center (CSCS), project s575.
\textbf{Author contributions: }Y.M. and K.B. conceived the project and planed the
research. Y.M. performed the transport measurements with the help of K.I. and analyzed the data.
K.A., A.M., and M.T. carried out the specific heat measurement. Y.A. synthesized the samples. A.S. performed
the theoretical calculations. Y.M., A.S., and K.B. wrote the manuscript.
All authors discussed the results and reviewed the manuscript.
\textbf{Competing interests: }The authors declare that they have no competing interests.
\textbf{Data and materials availability: }All data needed to evaluate the conclusions in the paper
are present in the paper and/or the Supplementary Materials. Additional data related to this paper
may be requested from the authors.

\clearpage

\begin{figure}[p]
\begin{center}
\includegraphics[width=12cm]{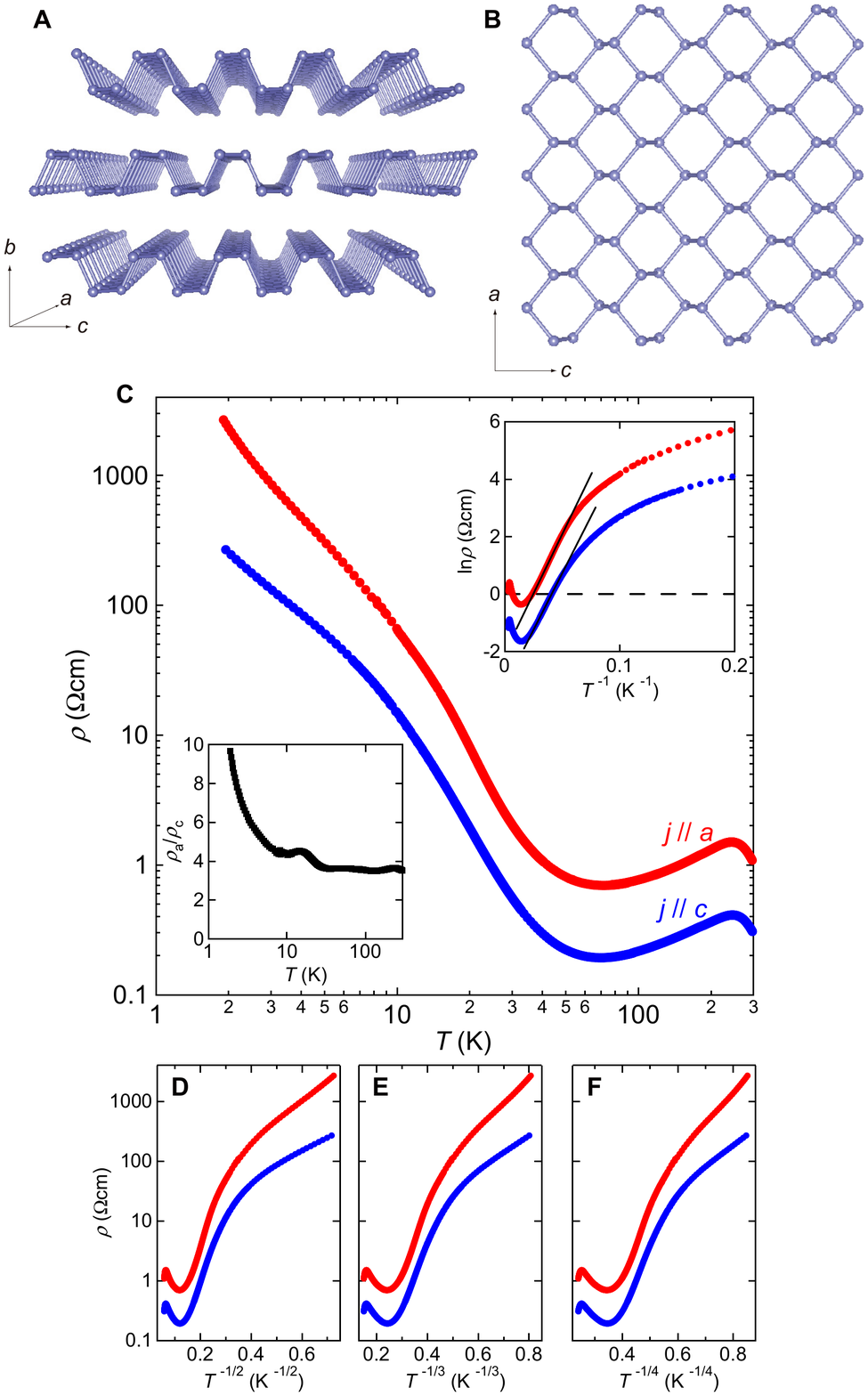}\\
\end{center}
{\bf Fig. 1. Crystal structure and resistivity.} ({\bf A}) Side view of the crystal structure of BP. ({\bf B}) Top view of single-layer BP. ({\bf C}) Temperature dependence of the electrical resistivity $\rho$ of BP along the $a$ and $c$ axes in a logarithmic scale. The upper inset shows an activated behavior of $\rho$ for $T>$ 20~K, where solid lines represent an Arrhenius behavior. The lower inset shows the ratio $\rho_{a}/\rho_{c}$, which quantifies charge transport anisotropy. At low temperatures,  variable range hopping governs, as shown in the lower panels where $\rho$ is plotted against $T^{-1/2}$ ({\bf D}), $T^{-1/3}$ ({\bf E}), and  $T^{-1/4}$({\bf F}).
\end{figure}

\begin{figure}[p]
\includegraphics[width=16cm]{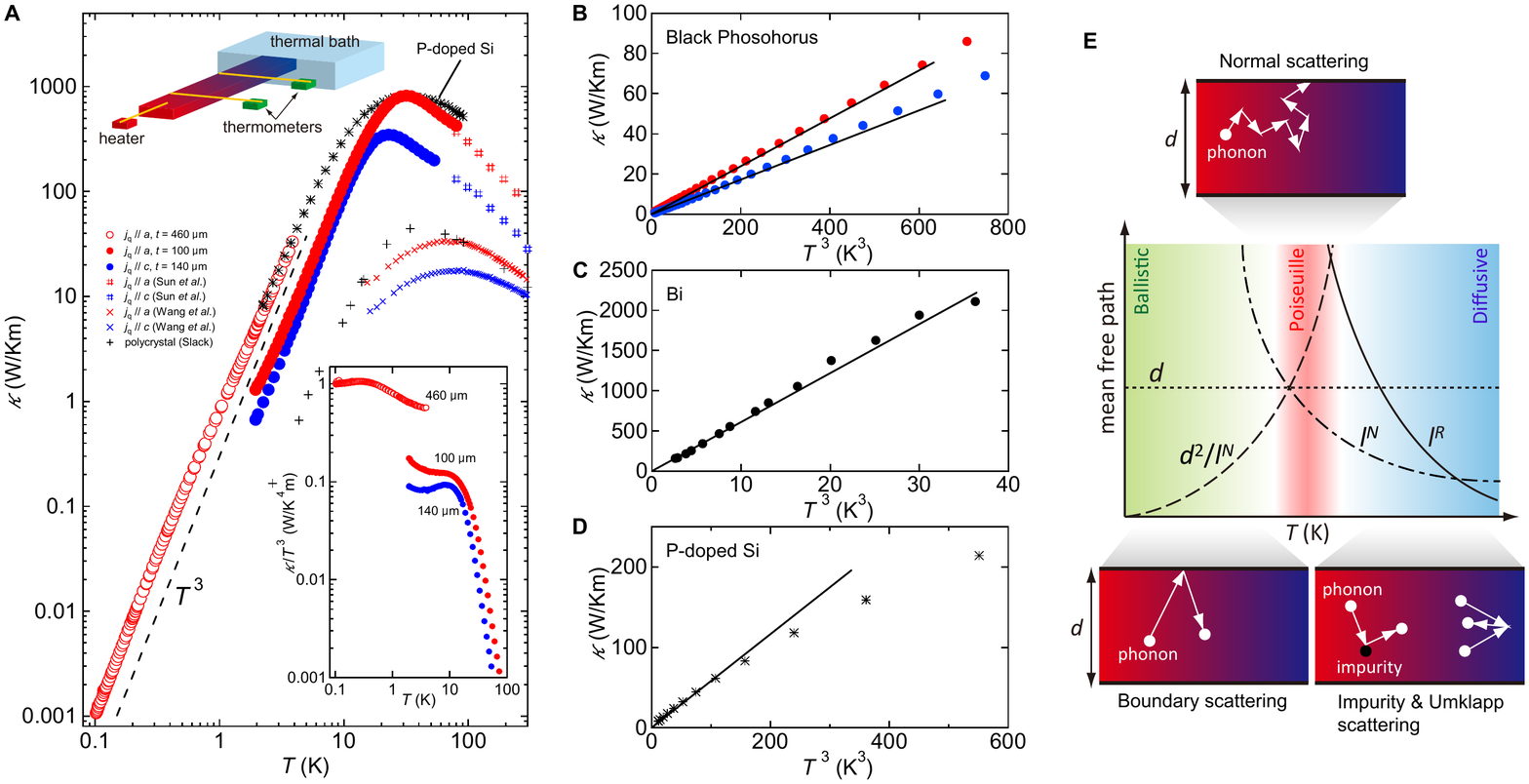}
\vspace*{-1.5cm}\\
{\bf Fig. 2. Thermal conductivity.} ({\bf A}) Temperature dependence of  the thermal conductivity along the two high-symmetry axes for the samples with a comparable thickness (solid circles). The $a$ axis thermal conductivity measurement is extended down to 0.1 K using the thicker sample (open circles). Our data is compared to three sets of previous reports~\cite{slack,wang,sun}. Thermal conductivity of P-doped Si obtained by using the same experimental setup is also shown.
Upper inset illustrates a schematic of
the measurement setup for the thermal conductivity.
Lower inset shows $\kappa$ divided by $T^3$ as a function of temperature. By plotting the thermal conductivity as a function of $T^{3}$ one can see how the ballistic regime, where $\kappa \propto$ $T^{3}$, evolves to the Poiseuille regime where $\kappa$ changes faster than $T^{3}$.
Such a behavior can be seen in BP ({\bf B}) and in the literature data of Bi~\cite{kopylov} ({\bf C}), but it is absent in P-doped Si ({\bf D}).
({\bf E}) Schematic representation of Poiseuille flow: the temperature dependence of mean-free-path of normal scattering $l^N$ (dashed-dotted line) and resistive scattering $l^R$ (solid line) and the sample dimension, $d$, in the three regimes of thermal transport. Poiseuille flow of phonons takes place in the limited temperature range between ballistic and diffusive transport when the inequality of $l^N\ll d\ll l^R$ is fulfilled. In the center of samples, the phonon mean-free-path (dashed line) becomes $l_{ph}\propto d^2/l^N$ in these conditions. The three small panels represent three distinct scattering mechanisms. Normal scattering between two phonons does not lead to any loss of momentum and does not degrade the heat current. In contrast, Umklapp scattering and collision with impurities lead to a loss in heat current and amplify thermal resistance. In any finite-size sample, phonons are also scattered by the boundaries.
\end{figure}

\begin{figure}[p]
\begin{center}
\includegraphics[width=15cm]{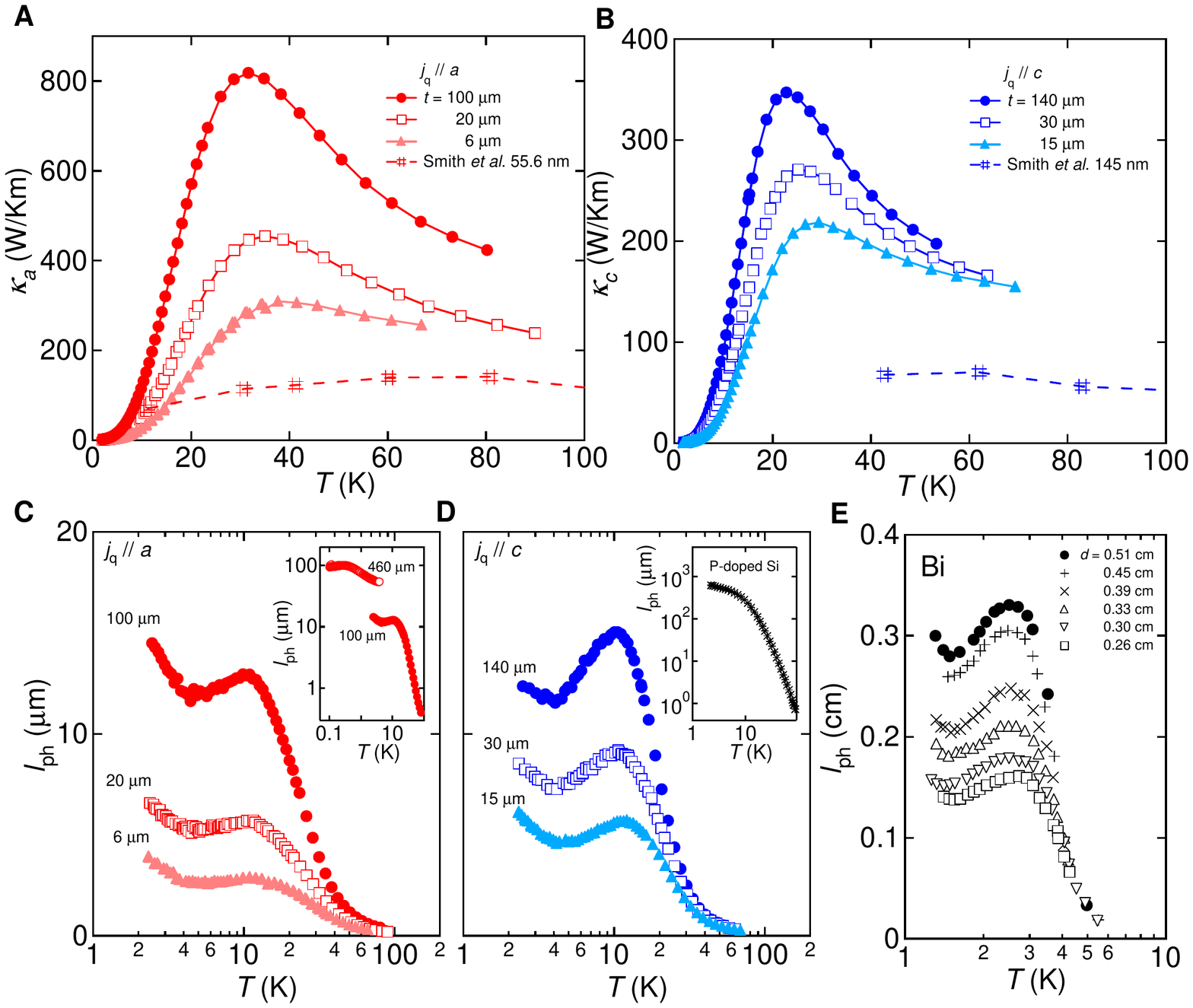}
\vspace{-7cm}
\end{center}
{\bf Fig. 3. Thickness dependence of thermal conductivity.} Temperature dependence of the thermal conductivity in samples with different thicknesses along the $a$ axis ({\bf A}) and along the $c$ axis ({\bf B}). The data from the thin BP flakes are also shown~\cite{smith}. Thermal conductivity shows a thickness dependence in the whole temperature range. The extracted phonon mean-free-path $l_{ph}$ in samples with different thicknesses along the $a$ axis ({\bf C}) and along the $c$ axis ({\bf D}). The local maximum and minimum are present in all samples. 
The insets of the panel {\bf C} and {\bf D} show plots of $l_{ph}$ vs $T$ in a logarithmic scale for BP including the thicker sample and for P-doped Si, respectively.
({\bf E}) The effective phonon mean-free-path of Bi for various average diameters from Ref.~\cite{kopylov}. 
\end{figure}

\begin{figure}[p]
\begin{center}
\includegraphics[width=16cm]{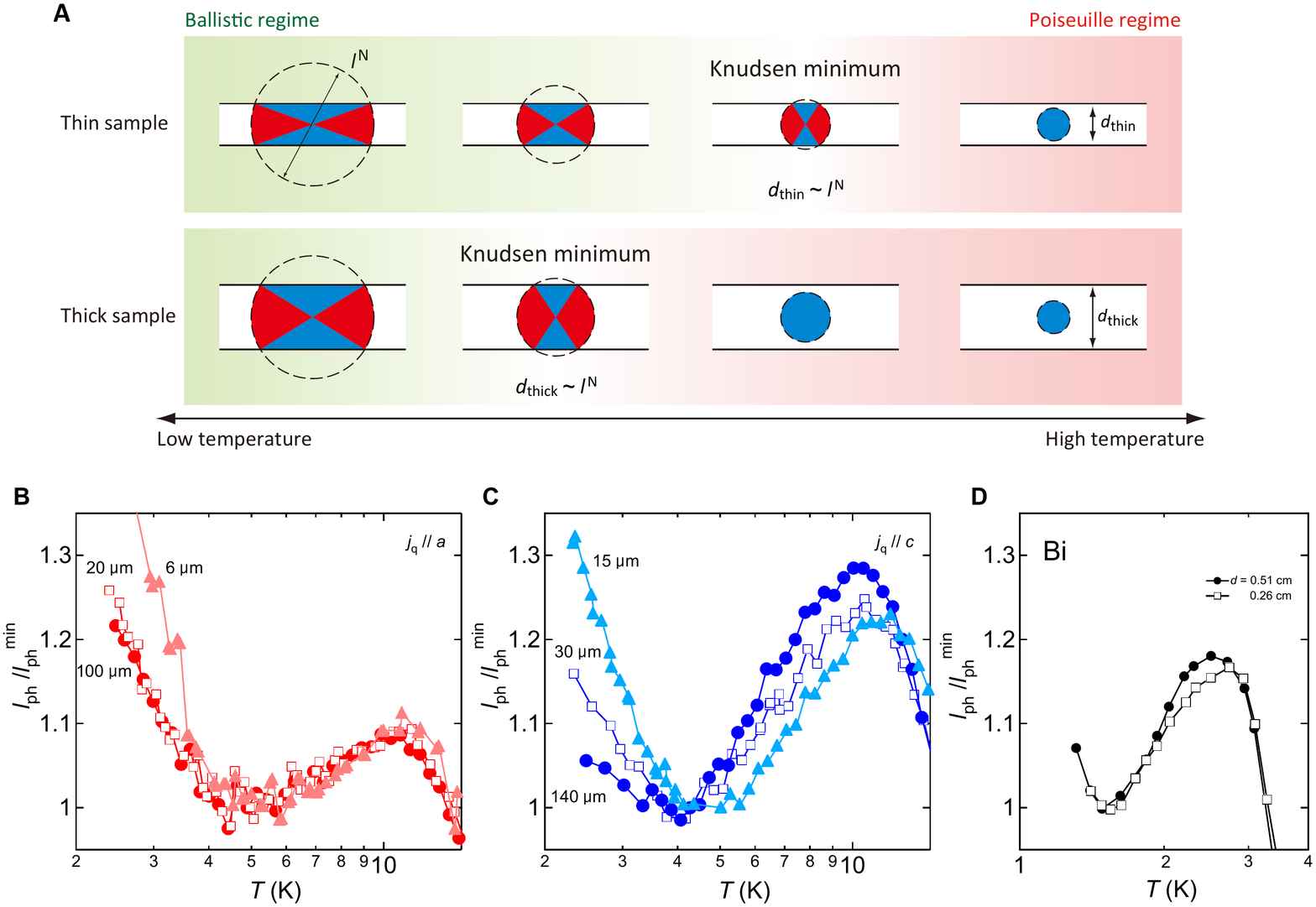}
\end{center}
{\bf Fig. 4. Thickness dependence of the Knudsen minimum.} ({\bf A}) Schematic illustration of the Knudsen minimum adapted from Ref.~\cite{ding}. The diffuse boundary scattering rate is effectively increased due to normal scattering when the mean-free-path of normal scattering $l^{N}$ represented by the dashed circle is comparable with the sample dimension $d$, producing a local minimum in the effective phonon mean-free-path. With increasing thickness, the minimum shifts to lower temperatures since the fraction of phonons which suffer numerous normal scattering (phonons in the red region) is larger in thicker sample. The phonon mean-free-path normalized by the value at the Knudsen minimum in samples with different thicknesses along the $a$ axis ({\bf B}) and the $c$ axis ({\bf C}), and Bi adapted from Ref.~\cite{kopylov} ({\bf D}).
\end{figure}

\begin{figure}[p]
\begin{center}
\includegraphics[width=16cm]{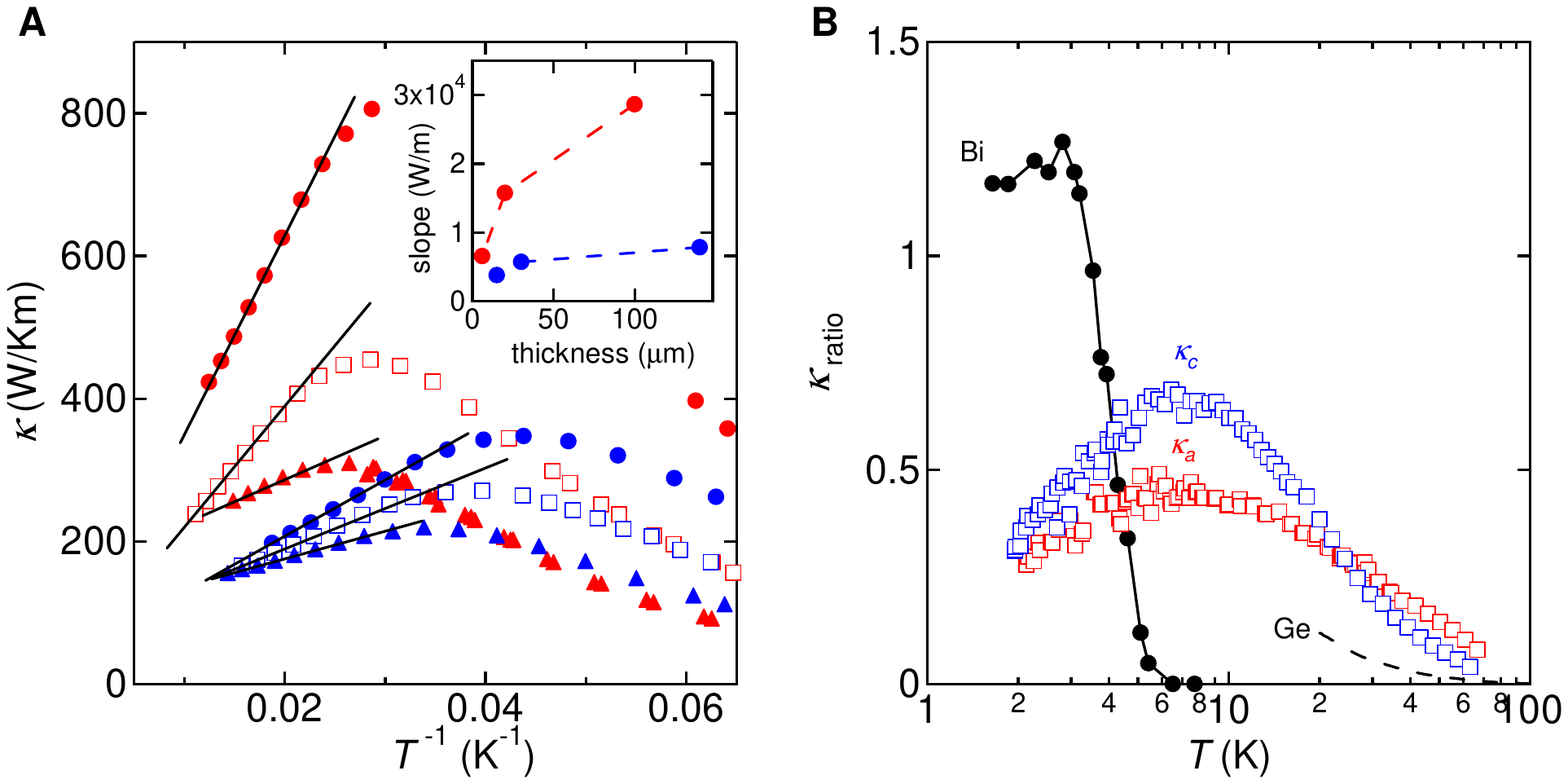}
\vspace{-13cm}\\
\end{center}
{\bf Fig. 5. $T^{-1}$ dependence of thermal conductivity and relative decrease of thermal conductivity with thickness.} ({\bf A}) $\kappa$ as a function of $T^{-1}$ in samples with different thicknesses. Note the gradual decrease in slope with thickness as shown in the inset. The temperature dependence of the relative decrease in $\kappa$, $\kappa_{\rm ratio}$, when the thickness decreases from 20(30) $\mu$m to 6(15) $\mu$m along the $a$($c$) axis ({\bf B}).
$\kappa_{\rm ratio}$ is defined as ($\kappa(t_{1})-\kappa(t_2)$)/$\kappa(t_2)$$\times$$t_2/(t_{1}-t_{2})$, where $t_{1,2}$ denotes different thickness. One expects this to become of the order of unity in the ballistic regime and zero in the high-temperature when heat transport is purely diffusive. 
These features are seen both in Bi~\cite{kopylov} (when the average diameter decreases from 0.51~cm to 0.26~cm) and in Ge~\cite{ziman} (when the width was decreased from 4~mm to 1~mm).
In BP samples of a typical size of 6~$\mu$m to 140~$\mu$m, $\kappa_{\rm ratio}$ persists (while slowly decreasing) in the diffusive regime.
\end{figure}

\begin{figure}[p]
\includegraphics[width=16cm]{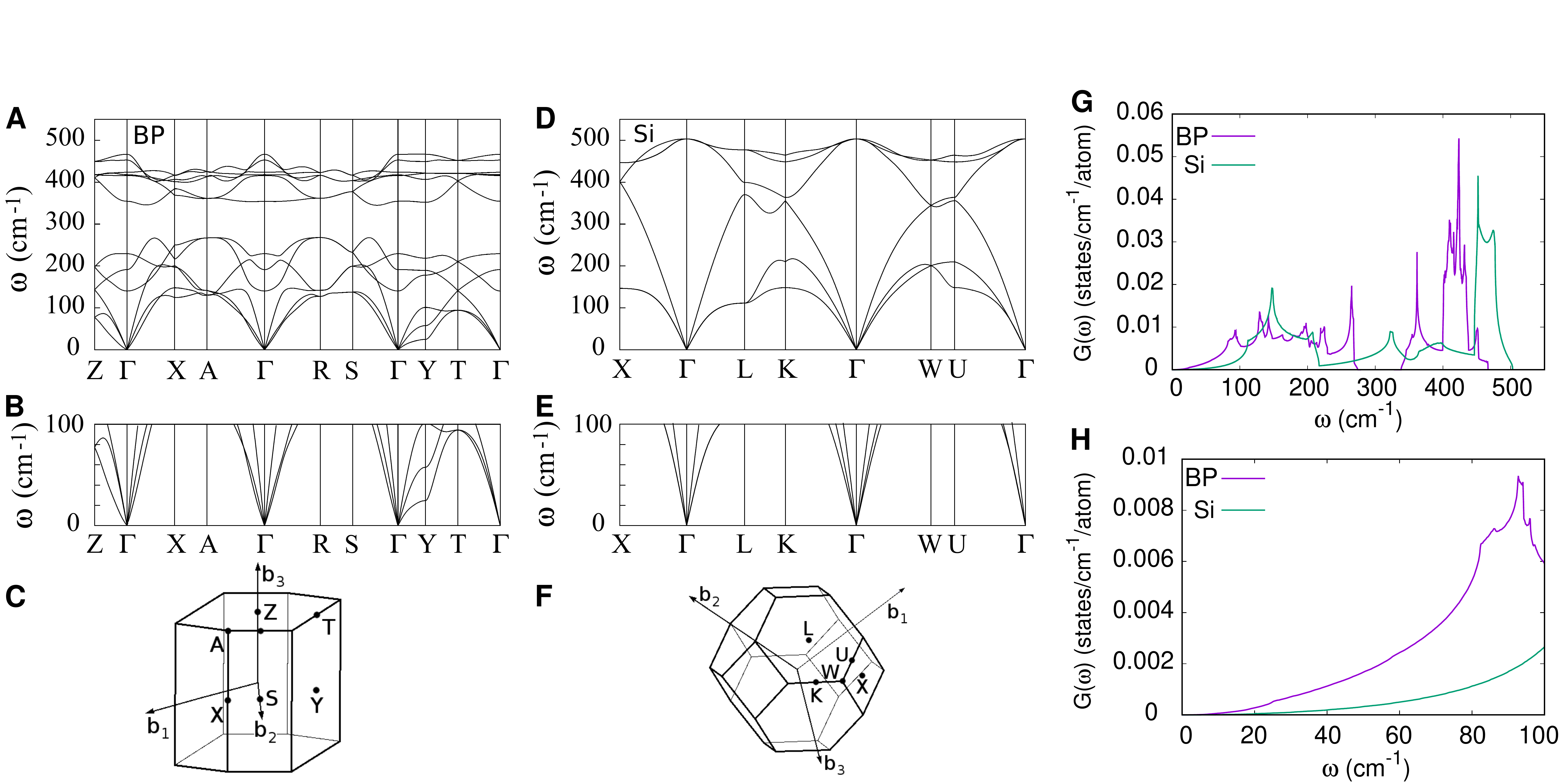}\\
\vspace*{-0cm}\\
{\bf Fig. 6. Phonon dispersions.} Calculated phonon dispersions of BP ({\bf A}) and Si ({\bf D}). Panels ({\bf B}) and ({\bf E}) show the blow-up in the region below 100 cm$^{-1}$. Panels ({\bf C}) and ({\bf F}) show the Brillouin zones along with the high-symmetry points. The out-of-plane direction in BP corresponds to the path along $\Gamma$-Y.
({\bf G}) Calculated phonon density of states (PDOS) of BP and Si. ({\bf H}) PDOS  below 100 cm$^{-1}$. The acoustic modes of BP are less dispersive than that of Si, leading to higher PDOS in BP in the low-frequency region. At low temperatures, BP should have much more phase space for the momentum-conserving phonon scattering processes among the linearly dispersive acoustic modes that are relevant for Poiseuille heat flow.
\end{figure}

\clearpage

\begin{center}

\vspace{0.3cm}

{\Large Supplementary Materials for\\[0.3cm]
\bfseries{Observation of Poiseuille Flow of Phonons in Black Phosphorus}}

\vspace{0.5cm}

{\large
Yo Machida,$^{1\ast}$ Alaska Subedi,$^{2,3}$ Kazuto Akiba,$^{4}$ Atsushi Miyake,$^{4}$\\ Masashi Tokunaga,$^{4}$
Yuichi Akahama,$^{5}$ Koichi Izawa,$^{1}$ and Kamran Behnia$^{6\ast}$\\
}

\vspace{0.4cm}

\normalsize{$^{1}$Department of Physics, Tokyo Institute of Technology, Meguro 152-8551, Japan}\\
\normalsize{$^{2}$Centre de Physique Th\'eorique, \'Ecole Polytechnique, CNRS, Universit\'e Paris-Saclay,}\\
\normalsize{F-91128 Palaiseau, France}\\
\normalsize{$^{3}$Coll\`ege de France, 11 place Marcelin Berthelot, 75005 Paris, France}\\
\normalsize{$^{4}$The Institute for Solid State Physics, University of Tokyo, Kashiwa 277-8581, Japan}\\
\normalsize{$^{5}$Graduate School of Material Science, University of Hyogo, Kamigori 678-1297, Japan}\\
\normalsize{$^{6}$Laboratoire Physique et Etude de Mat\'{e}riaux (CNRS-UPMC), ESPCI Paris,}\\
\normalsize{PSL Research University, 75005 Paris, France}\\
\vspace{0.5cm}
\normalsize{$^\ast$To whom correspondence should be addressed;}\\
\normalsize{E-mail:  machida@ap.titech.ac.jp; kamran.behnia@espci.fr.}

\end{center}

\vspace{0.5cm}

\noindent
Materials and Methods\\
Supplementary Text\\
Figs.~S1 to S4\\
Tables~S1 to S2\\
References \textit{(1-14)}

\clearpage

\section*{Materials and Methods}
\subsection*{Samples}
Single crystals of  black phosphorus (BP) were synthesized under high pressure~\cite{endo}. 
Magneto-phonon resonance which  requires reasonable purity was observed in a sample from the batch used in the present experiments~\cite{akiba}.
A single crystal of P-doped Si was provided by Institute of Electronic Materials Technology (Warsaw).
The electrical resistivity and the thermal conductivity measurements were performed along the $a$ and $c$ axes of BP. Each sample has a rectangular shape with edges parallel to the three high-symmetry axes; the $a$, $b$, and $c$ axes. Length of the samples are summarized in Table S1. The $a$($c$) axis sample has the largest length along the $a$($c$) axis, respectively. 
The thickness dependence of the thermal conductivity was investigated by using the 
same sample.
The thickness (number of layers) along the $b$ axis was decreased during the course of the investigations by cleaving.
In Table S1, number after the hyphen of the sample label represents a sequence of the experiments from the
thickest to the thinnest sample. 
The resistivity data for the sample $a$1-1 and $c$1-1 are displayed in Fig. 1 of the main text. 
The thermal conductivity data for the sample $a0$, $a$1-1, and $c$1-1 are displayed in Fig. 2 of the main text. The thermal conductivity of the sample $a$1-1, $a$1-2, and $a$1-3 and those of  $c$1-1, $c$1-2, and $c$1-3 are shown in Figs. 3-5.
\begin{table}[h]
{\bf Table S1. Sample dimensions.} Length (in $\mu$m) of the samples along the three high-symmetry axes.\\
  \begin{center}
    \begin{tabular}{l|c|ccc|cc|ccc} \hline
      Sample label &a0&$a$1-1&$a$1-2&$a$1-3&$a$2-1&$a$2-2&$c$1-1&$c$1-2&$c$1-3\\ \hline \hline
      Along $a$-axis &3300& 2400 & 2400 & 2400 & 2700 & 2700 & 200 & 200 & 200\\
      Along $c$-axis &900& 320 & 320 & 320 & 160 & 160 & 2200 & 2200 & 2200\\
      Along $b$-axis &460& 100 & 20 & 6 & 120 & 8 & 140 & 30 & 15\\  \hline
    \end{tabular}
  \end{center}
\end{table}

\subsection*{Heat and electrical transport measurements}
The thermal conductivity was measured by using a home build system. We employed a standard one-heater-two-thermometers steady-state method. The measurements were performed in the temperature range between 0.1 K and 80 K.
The thermometers, the heater and the cold finger were connected to the sample by gold wires of 25 $\mu$m diameter. The
gold wires were attached on the BP sample by Dupont 4922N silver paste and were soldered by indium on the P-doped Si sample. The contact resistances were less than 10 $\Omega$ at room temperature. The temperature difference generated in the sample by the heater was determined by measuring the local temperature with two thermometers (Cernox resistors in the $^4$He cryostat and RuO$_2$ resistors in the dilution refrigerator). The heat loss along manganin wires connected to the two thermometers and the heater are many orders of magnitude lower than the thermal current along the sample including the thinner ones (Fig.~S1)~\cite{zavaritskii,corzett,peroni}. The heat loss by radiation is negligible in the temperature range of our study ($T<$ 80~K), since it follows a $T^4$ behavior. 
The heat loss by residual gas is also negligible because the measurements were carried out in an evacuated 
chamber with a vacuum level better than $10^{-4}$ Pa. Moreover, the external surface of the chamber is directly in the helium bath at 4.2~K, so
that residual gas is condensed on the cold wall.
The experimental uncertainty in the thermal conductivity arising from the measurements of the thermometer resistances
is less than 1 $\%$ in the whole temperature range.
The main source of uncertainty results from uncertainty in the measured thickness of the samples, which is about 5 $\%$ in the 
thinnest sample.
The increase of thermal conductivity with the sample thickness, however, dominates over the experimental uncertainty.

The same contacts and wires were used for the electrical resistivity measurements by a four-point technique. A DC electric current was applied along the sample using the manganin wire attached on the heater. The Keithley 2182A nanovoltmeter was employed for the measurement of electrical voltage. The input impedance of the nanovoltmeter is larger than 10 G$\Omega$, which is well above the resistance of the sample even at the lowest temperatures ($R\sim$ 5~M$\Omega$).

\subsection*{Computational}
The phonon dispersions were calculated using density functional perturbation theory~\cite{dfpt} as implemented in the {\sc quantum espresso} package~\cite{qe}. We used the generalized gradient approximation~\cite{pbe} and the pseudopotentials generated by Garrity \textit{et al.}~\cite{gbrv}. For  BP, additional van der Waals correction using Grimme's semiempirical recipe was applied~\cite{grimme,baro09}. Planewave cutoffs of 50 and 250 Ry were used for the basis-set and charge density expansions, respectively. A $12\times12\times12$ grid was used for the Brillouin zone integration in the self-consistent density functional theory calculations. A Marzari-Vanderbilt smearing of 0.02 Ry was used in the calculations for BP. The dynamical matrices were calculated on an $8\times8\times8$ grid, and the phonon dispersions and density of states were obtained by Fourier interpolation.

\section*{Supplementary Text}
\subsection*{Reproducibility of thermal conductivity}
To check the reproducibility of the thermal conductivity and its thickness dependence, we redid the thermal conductivity measurements
along the $a$ axis using the sample $a$2. Figure S2 shows temperature dependence of the thermal conductivity $\kappa_a$ for the sample $a$1-1, $a$1-2, and $a$1-3 together with those for the sample $a$2-1 and $a$2-2. The figure clearly indicates that for the sample with the comparable thickness, the magnitude and the temperature variation of the thermal conductivity are well reproduced. This
eliminates artificial origin of the thickness dependence such as difference in impurity concentrations and crystal imperfections between the individual samples.

\subsection*{Specific heat}
Specific heat $C$ of BP was measured between 2 K and 100 K using a relaxation method of the heat capacity option in a Quantum Design, PPMS instrument. A single-crystalline BP was used for the measurement. 
Our specific heat data agree well with the previous report~\cite{paukov} as shown in Fig.~S3A. Below 5 K, the specific heat follows a $T^3$ behavior and is fitted to an expression $C=\beta T^3$ with $\beta$ = 6.7 $\times$ 10$^{-5}$ J/molK$^4$, as expected from the Debye model (Fig.~S3B). 
The Debye temperature $\Theta_D$ estimated from $\beta$ is $\Theta_D$ = 306 K, which is consistent with 
those obtained from the theoretical calculation~\cite{kaneta}.
The calculated specific heat based on the phonon dispersions from the first principles (Fig.~6A) reasonably reproduces not only the $T^3$ dependence but also matches the magnitude of the measured specific heat within a margin of 25 percent as shown in the inset of Fig.~S3B.

\subsection*{Phonon velocity}
The phonon velocity $v$ of BP in each acoustic mode along the $c$, $a$, and $b$ axes determined from the initial slope of phonon dispersion near the $\Gamma$ point is listed in Table~S2.
LA and LA represent longitudinal and transverse modes, respectively. 
Our calculated values are comparable to those obtained from the inelastic neutron scattering experiments~\cite{fujii} and the previous theoretical calculation~\cite{kaneta}.
We note that BP has small anisotropy of the phonon velocity between 
in-plane direction ($ac$ plane) and out-of-plane direction ($b$ axis) in spite of its layered structure (see Fig.~S4).
The specific heat data shown in Fig.~S3A and the mean phonon velocity along the $a$ and $c$ axes, $\langle v_a\rangle$ = 0.536$\times$10$^4$ m/s and $\langle v_c\rangle$ = 0.354$\times$10$^4$ m/s,
are used to evaluate the effective phonon mean-free-path $l_{ph}$ shown in Fig.~3 and Fig.~4 in the main text.

\begin{table}[h]
{\bf Table S2. Phonon velocity of black phosphors.} Phonon velocity (in 10$^4$ m/s) along different orientations obtained in the present study is compared with those from the neutron scattering studies~\cite{fujii} and the previous theoretical calculation~\cite{kaneta}. 
Notice the consistency and the low anisotropy.\\
  \begin{center}
    \begin{tabular}{cc|c|c|c} \hline
      & & This work & Exp. & Calc. \\
      & & & Fujii $et al$.~\cite{fujii} & Kaneta $et al$.~\cite{kaneta} \\ \hline\hline
      $ $&LA$_c$&0.454&0.46&0.436\\
      $v_c$&TA$_a$&0.490&0.46&0.419\\
      $ $&TA$_b$&0.119&-&0.109\\\cline{2-5}
      $ $&Average&0.354&-&0.321\\\hline
      $ $&LA$_a$&0.833&0.96&0.954\\
      $v_a$&TA$_c$&0.488&-&0.419\\
      $ $&TA$_b$&0.287&-&0.254\\\cline{2-5}
      $ $&Average&0.536&-&0.542\\\hline
      $ $&LA$_b$&0.505&0.510&0.442\\
      $v_b$&TA$_c$&0.160&0.170&0.109\\
      $ $&TA$_a$&0.289&0.310&0.254\\\cline{2-5}
      $ $&Average&0.318&0.330&0.268\\\hline

    \end{tabular}
  \end{center}
\end{table}

\clearpage

\begin{figure}
\begin{center}
\includegraphics[width=15cm]{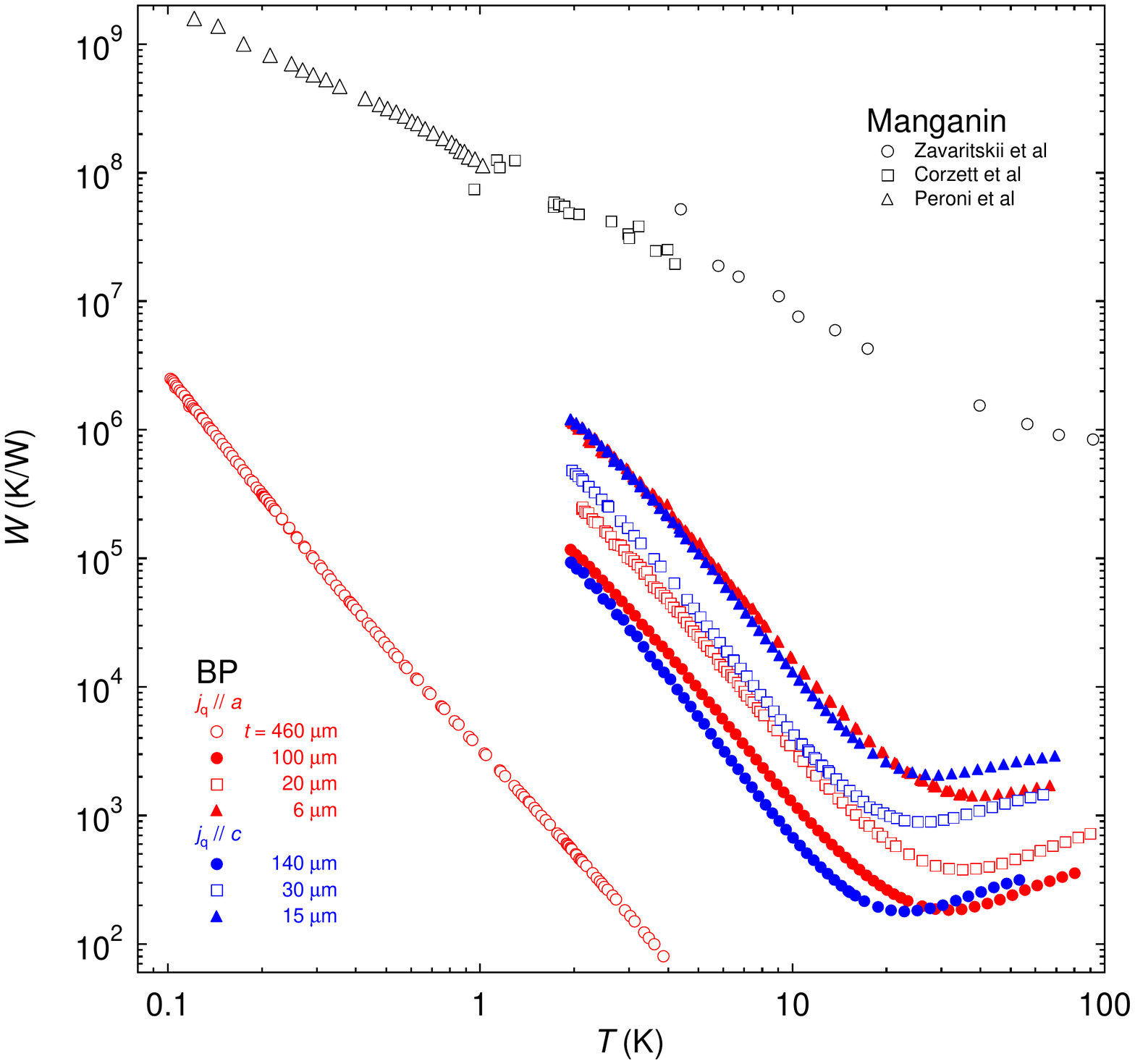}\\
\end{center}
{\bf Fig. S1. Thermal resistance.} Thermal resistance $W$ of the BP samples and the manganin wires connected to the thermometers and heaters~\cite{zavaritskii,corzett,peroni}.
\end{figure}

\begin{figure}
\begin{center}
\includegraphics[width=15cm]{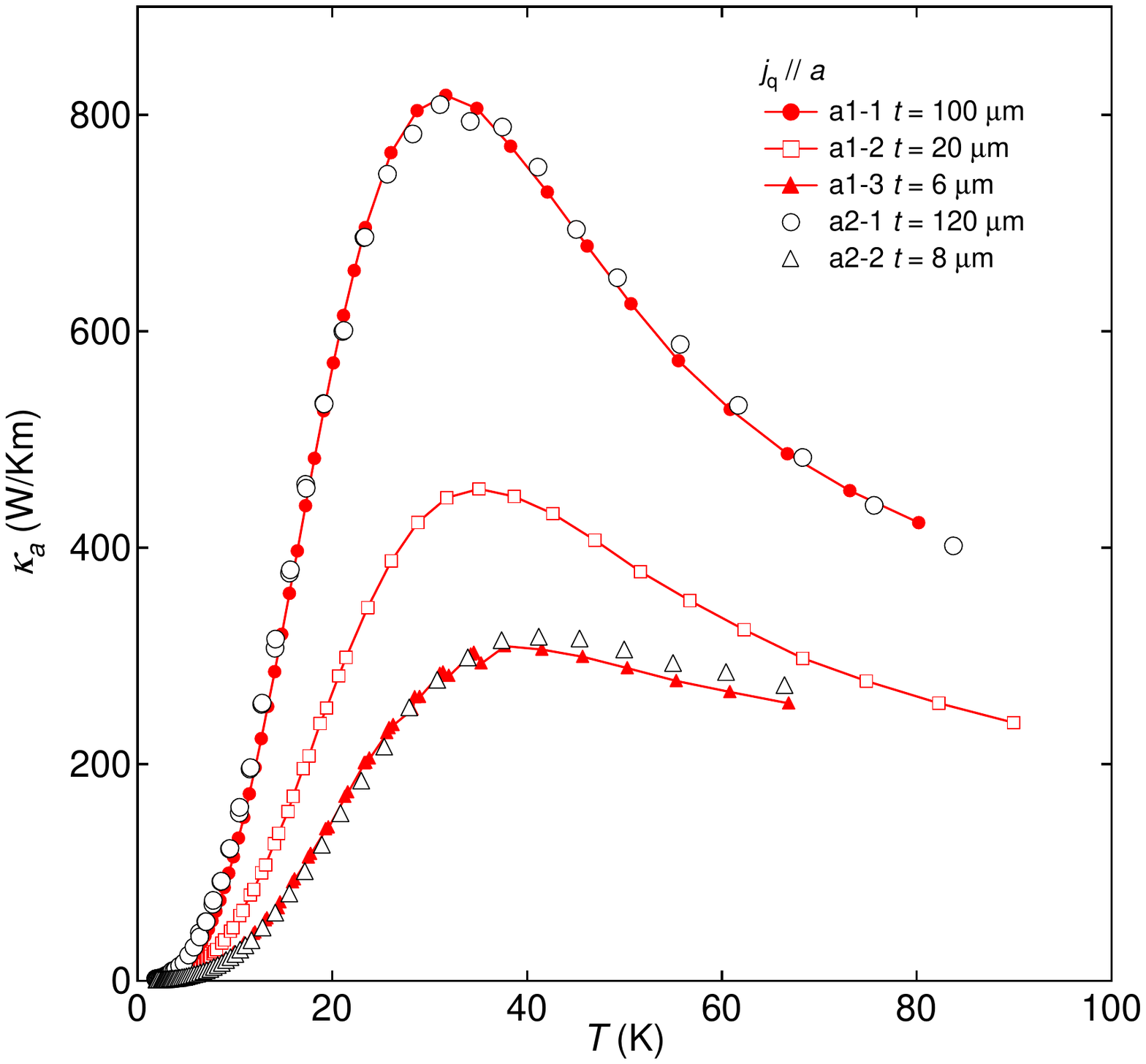}\\
\end{center}
{\bf Fig. S2. Reproducibility of thermal conductivity.} Temperature dependence of the thermal conductivity $\kappa_a$ along the $a$ axis for the sample $a$1-1, $a$1-2, and $a$1-3 denoted by red symbols,
together with those for $a$2-1 and $a$2-2 denoted by black symbols.
\end{figure}

\begin{figure}
\begin{center}
\includegraphics[width=17cm]{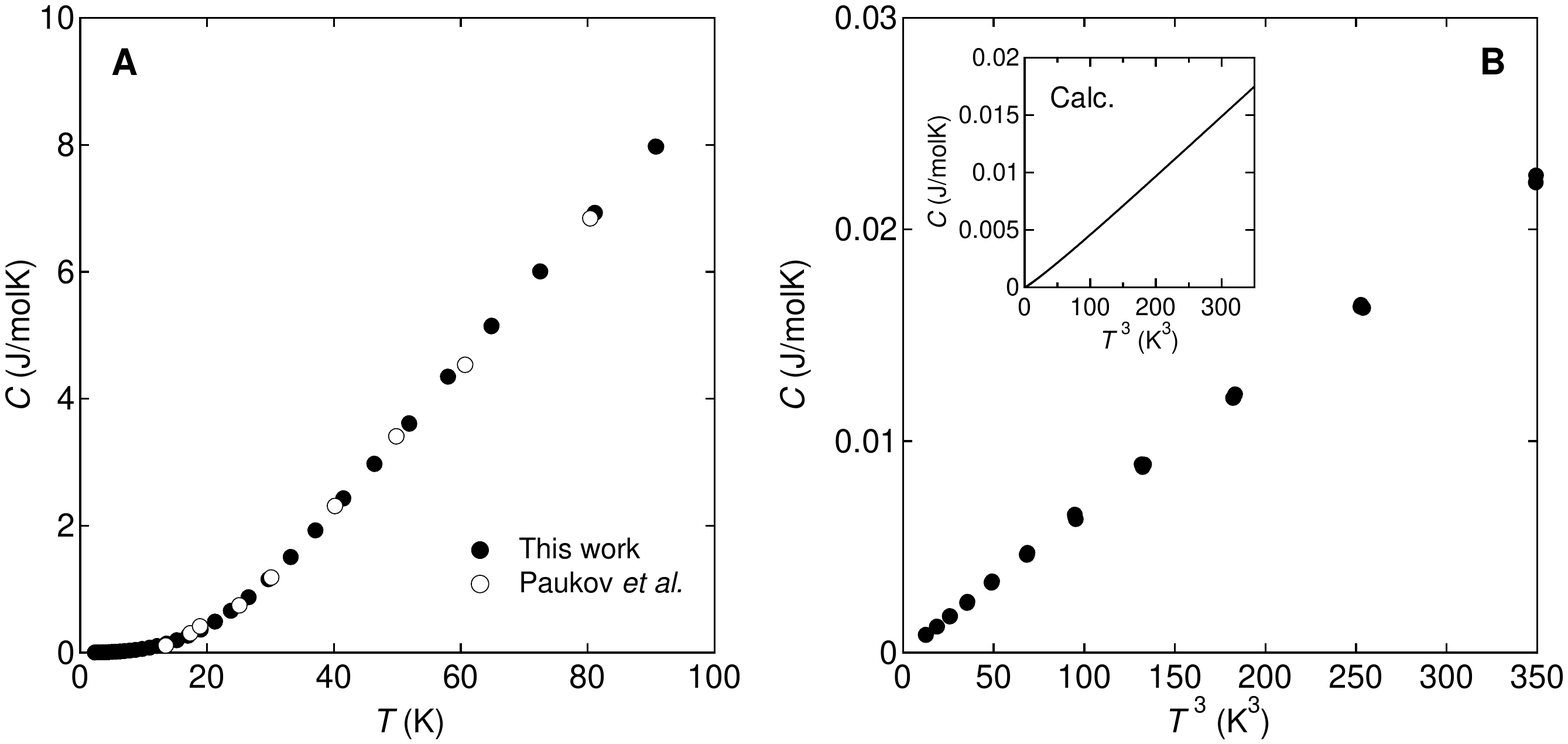}
\end{center}
{\bf Fig. S3. Specific heat.} ({\bf A}) Specific heat $C$ of BP as a function of temperature. The specific heat data adapted from Ref.~\cite{paukov} is also shown (open circles). ({\bf B}) Experimental (main panel) and 
theoretical (inset) specific heat as a function of $T^3$. The theoretical slope is 0.8 times the experimental one.
\end{figure}

\begin{figure}
\begin{center}
\includegraphics[width=13cm]{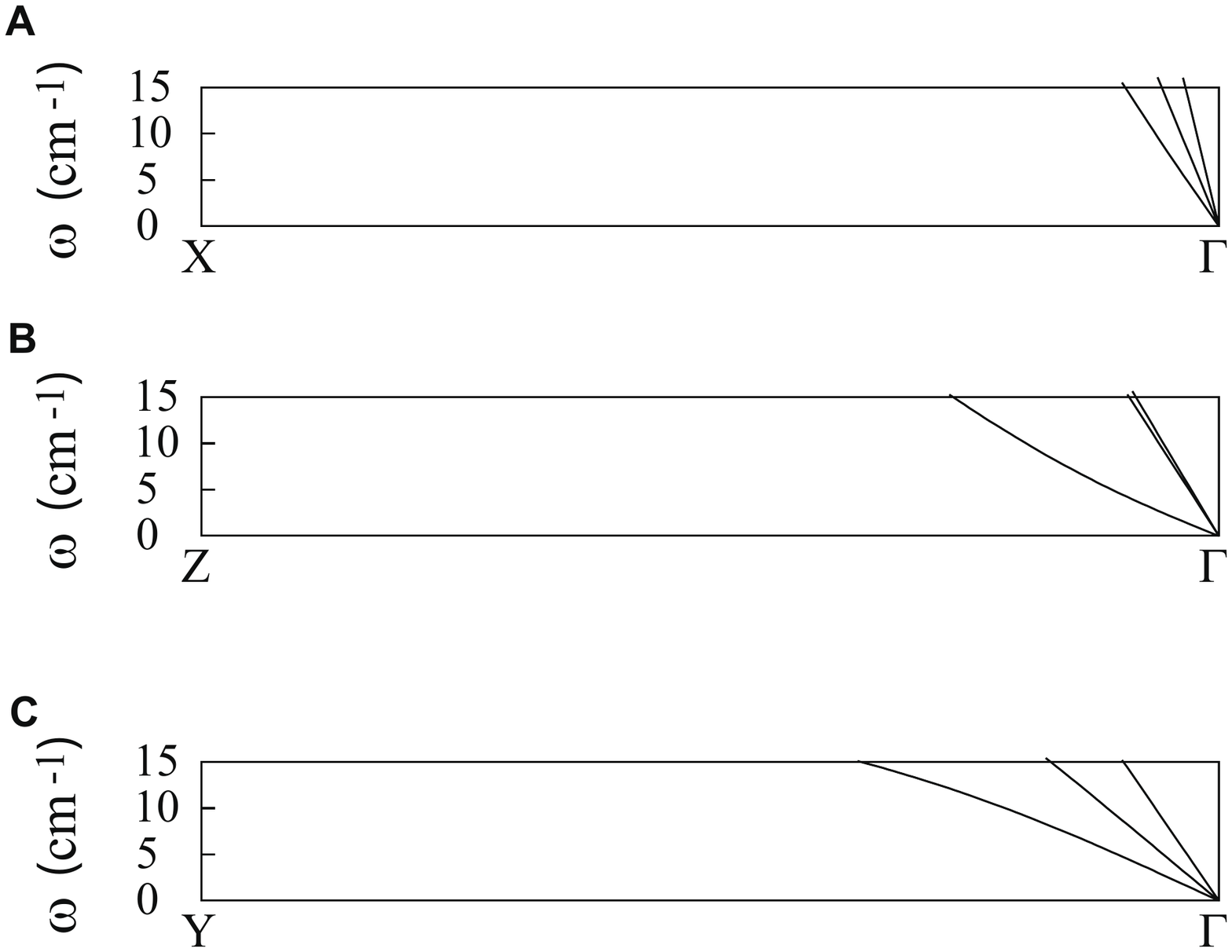}
\vspace{-4cm}
\end{center}
{\bf Fig. S4. Phonon dispersions.} The phonon dispersions of BP zoomed to 15 cm$^{-1}$ are shown along the high-symmetry directions, $\Gamma$-X ($a$ axis) ({\bf A}), $\Gamma$-Z ($c$ axis) ({\bf B}), and $\Gamma$-Y ($b$ axis) ({\bf C}).
\end{figure}


\begin{thebibliography}{99}
\bibitem{gurzhi} R. N. Gurzhi, Hydrodynamic effects in solids at low temperature, Sov. Phys.-Usp. \textbf{11}, 255 (1968).
\bibitem{xia} F. Xia, H. Wang, and Y. Jia, Rediscovering black phosphorus as an anisotropic layered material for optoelectronics and electronics, Nat. Commun. \textbf{5}, 4458 (2014).
\bibitem{ling} X. Ling, H. Wang, S. Huang, F. Xia, and M. S. Dresselhaus, The renaissance of black phosphorus, Proc. Natl. Acad. Sci. USA \textbf{112}, 4523 (2015).
\bibitem{das} S. Das, W. Zhang, M. Demarteau, A. Hoffmann, M. Dubey, and A. Roelofs, Tunable transport gap in phosphorene, Nano Lett. \textbf{14}, 5733 (2014).
\bibitem{fei} R. Fei, A. Faghaninia, R. Soklaski, J.-A. Yan, C. Lo, and L. Yang, Enhanced thermoelectric efficiency via orthogonal electrical and
thermal conductances in phosphorene, Nano Lett. \textbf{14}, 6393 (2014).
\bibitem{akahama} Y. Akahama, S. Endo, and S. Narita, Electrical properties of black phosphorus single crystal, J. Phys. Soc. Jpn. \textbf{52}, 2148 (1983).
\bibitem{akiba} K. Akiba, A. Miyake, Y. Akahama, K. Matsubayashi, Y. Uwatoko, H. Arai, Y. Fuseya, and M. Tokunaga, Anomalous quantum transport properties in semimetallic black phosphorus, J. Phys. Soc. Jpn. \textbf{84}, 073708 (2015).
\bibitem{xiang} Z. J. Xiang, G. J. Ye, C. Shang, B. Lei, N. Z. Wang, K. S. Yang, D. Y. Liu, F. B. Meng, X. G. Luo, L. J. Zou, Z. Sun, Y. Zhang, and X. H. Chen, Pressure-induced electronic transition in black phosphorus, Phys. Rev. Lett. \textbf{115}, 186403 (2015).
\bibitem{akibaHall} K. Akiba, A. Miyake, Y. Akahama, K. Matsubayashi, Y. Uwatoko, and M. Tokunaga, Two-carrier analyses of the transport properties of black phosphorus under pressure, Phys. Rev. B \textbf{95}, 115126 (2017).
\bibitem{slack} G. A. Slack, Thermal conductivity of elements with complex lattices: {B}, {P}, {S}, Phys. Rev. \textbf{139}, A507 (1965).
\bibitem{flores} E. Flores, J. R. Ares, A. Castellanos-Gomez, M. Barawi, I. J. Ferrer, and C. S$\acute{\rm a}$nchez, Thermoelectric power of bulk black-phosphorus, Appl. Phys. Lett. \textbf{106}, 022102 (2015).
\bibitem{wang} Y. Wang, G. Xu, Z. Hou, B. Yang, X. Zhang, E. Liu, X. Xi, Z. Liu, Z. Zeng, W. Wang, and G. Wu, Large anisotropic thermal transport properties observed in bulk single crystal black phosphorus, Appl. Phys. Lett. \textbf{108}, 092102 (2016).
\bibitem{sun} B. Sun, X. Gu, Q. Zeng, X. Huang, Y. Yan, Z. Liu, R. Yang, and Y. K. Koh, Temperature dependence of anisotropic thermal conductivity tensor of bulk black phosphorus, Adv. Mater. \textbf{29}, 1603297 (2016).
\bibitem{smith} B. Smith, B. Vermeersch, J. Carrete, E. Ou, J. Kim, N. Mingo, D. Akinwande, and L. Shi, Temperature and thickness dependences of the anisotropic in-plane thermal conductivity of black phosphorus, Adv. Mater. \textbf{29}, 1603756 (2017).
\bibitem{hu} S. Hu, J. Xiang, M. Lv, J. Zhang, H. Zhao, C. Li, G. Chen, W. Wang, and P. Sun, Intrinsic and extrinsic electrical and thermal transport of bulk black phosphorus, Phys. Rev. B \textbf{97}, 045209 (2018).
\bibitem{beck} H. Beck, P. F. Meier, and A. Thellung, Phonon hydrodynamics in solids, Phys. Stat. Sol. (a) \textbf{24}, 11 (1974).
\bibitem{thacher} P. D. Thacher, Effect of boundaries and isotopes on the thermal conductivity of {L}i{F}, Phys. Rev. \textbf{156}, 975 (1967).
\bibitem{kaneta} C. Kaneta, H. Katayama-Yoshida, and A. Morita, Lattice dynamics of black phosphorus, Solid State Commun. \textbf{44}, 613 (1982).
\bibitem{fujii} Y. Fujii, Y. Akamaha, S. Endo, S. Narita, and Y. Yamada, Inelastic neutron scattering study of acoustic phonons of black phosphorus, Solid State Commun. \textbf{44}, 579 (1982).
\bibitem{kopylov} V. N. Kopylov and L. P. Mezhov-Deglin, Investigation of the kinetic coefficients of bismuth at helium temperatures, Zh. Eksp. Teor. Fiz. \textbf{65}, 720 (1973).
\bibitem{deglin} L. P. Mezhov-Deglin, Measurement of the thermal conductivity of crystalline {H}e$^4$, Zh. Eksp. Teor. Fiz. \textbf{49}, 66 (1965).
\bibitem{thomlinson} W. C. Thomlinson, Evidence for anomalous phonon excitations in solid {H}e$^3$, Phys. Rev. Lett. \textbf{23}, 1330 (1969).
\bibitem{zholonko} N. N. Zholonko, Poiseuille flow of phonons in solid hydrogen, Phys. Solid State \textbf{48}, 1678 (2006).
\bibitem{glass} C. J. Glassbrenner and G. A. Slack, Thermal conductivity of silicon and germanium from 3~K to the melting point, Phys. Rev. \textbf{134}, A1058 (1964).
\bibitem{slackSi} G. A. Slack, Thermal conductivity of pure and impure silicon, silicon carbide, and diamond, J. Appl. Phys. \textbf{35}, 3460 (1964).
\bibitem{deglinBi} L. P. Mezhov-Deglin, V. N. Kopylov, and $\acute{\rm E}$. S. Medvedev, Contributions of various phonon relaxation mechanisms to the thermal resistance of the crystal lattice of bismuth at temperatures below 2~{K}, Zh. Eksp. Teor. Fiz. \textbf{67}, 1123 (1974).
\bibitem{rogers} S. J. Rogers, Transport of heat and approach to second sound in some isotopically pure alkali-halide crystals, Phys. Rev. B \textbf{3}, 1440 (1971).
\bibitem{cepellotti} A. Cepellotti, G. Fugallo, L. Paulatto, M. Lazzeri, F. Mauri, and N. Marzaria, Phonon hydrodynamics in two-dimensional materials, Nat. Commun. \textbf{6}, 6400 (2015).
\bibitem{narayanamurti}  V. Narayanamurti and R. C. Dynes, Observation of second sound in bismuth, Phys. Rev. Lett. \textbf{28}, 1461 (1972).
\bibitem{moll} P. J. W. Moll, P. Kushwaha, N. Nandi, B. Schmidt, and A. P. Mackenzie, Evidence for hydrodynamic electron flow in PdCoO$_{2}$, Science  \textbf{351}, 1061 (2016).
\bibitem{eu} B. C. Eu, Generalization of the Hagen-Poiseuille velocity profile to non-Newtonian fluids and measurement of their viscosity,  Am. J. Phys. \textbf{58}, 83 (1990).
\bibitem{ding} Z. Ding, J. Zhou, B. Song, V. Chiloyan, M. Li, T-H. Liu, and G. Chen, Phonon hydrodynamic heat conduction and Knudsen minimum in graphite, Nano Lett.  \textbf{18}, 638 (2018).
\bibitem{broido} D. A. Broido, M. Malorny and G. Birner, N. Mingo, and D. A. Stewart, Intrinsic lattice thermal conductivity of semiconductors from first principles, Appl. Phys. Lett. \textbf{91}, 231922 (2007).
\bibitem{ziman} J. Ziman, \textit{Electrons and Phonons: The Theory of Transport Phenomena in Solids} (Oxford University Press, 2001).
\bibitem{fugallo} G. Fugallo, A. Cepellotti, L. Paulatto, M. Lazzeri, N. Marzari, and F. Mauri, Thermal conductivity of graphene and graphite: Collective
excitations and mean free paths, Nano Lett.  \textbf{14}, 6109 (2014).
\bibitem{leeHydro} S. Lee, D. Broido, K. Esfarjani, and G. Chen, Hydrodynamic phonon transport in suspended graphene, Nat. Commun. \textbf{6}, 6290 (2015).
\bibitem{zhan16} J. Zhang, H. J. Liu, L. Cheng, J. Wei, J. H. Liang, D. D. Fan, P. H. Jiang, L. Sun, and J. Shi, High thermoelectric performance can be achieved in black phosphorus, J. Mater. Chem. C {\bf 4}, 991 (2016).
\bibitem{gian91} P. Giannozzi, S. de Gironcoli, P. Pavone, and S. Baroni, {\it Ab initio} calculation of phonon dispersions in semiconductors, Phys. Rev. B {\bf 43}, 7231 (1991).
\bibitem{footnote} V. Martelli, J. L. Jim$\acute{\rm e}$nez, M. Continentino, E. Baggio-Saitovitch, and K. Behnia, Phys. Rev. Lett. \textbf{120}, 125901 (2018).
\bibitem{littlewood} P. Littlewood, The crystal structure of IV-VI compounds. I. Classification and description, J. Phys. C Solid State Phys. \textbf{13}, 4855 (1980).
\bibitem{behnia} K. Behnia, Finding merit in dividing neighbors, Science \textbf{351}, 124 (2016).
\bibitem{endo} S. Endo, Y. Akahama, S. Terada, and S. Narita, Growth of large single crystals of black phosphorus under high pressure, J. Appl. Phys. \textbf{21}, L482 (1982).
\bibitem{nakamae} S. Nakamae, K. Behnia, L. Balicas, F. Rullier-Albenque, H. Berger, and T. Tamegai, Effect of controlled disorder on quasiparticle heat transport in Bi2212, Phys. Rev. B \textbf{63}, 184509 (2001).
\bibitem{UPt3} Y. Machida, A. Itoh, Y. So, K. Izawa, Y. Haga, E. Yamamoto, N. Kimura, Y. Onuki, Y. Tsutsumi, and K. Machida, Twofold Spontaneous Symmetry Breaking in the Heavy-Fermion Superconductor UPt$_3$, Phys. Rev. Lett. \textbf{108}, 157002 (2012).
\bibitem{YRS} Y. Machida, K. Tomokuni, K. Izawa, G. Lapertot, G. Knebel, J.-P. Brison, and J. Flouquet, Verification of the Wiedemann-Franz Law in YbRh$_2$Si$_2$ at a Quantum Critical Point, Phys. Rev. Lett. \textbf{110}, 236402 (2013).
\bibitem{zavaritskii} N. V. Zavaritskii and A. G. Zeldovich,  Thermal conductivity of some technical materials at low temperatures, Zhur. Tekh. Fiz. \textbf{26}, 2032, (1956).
\bibitem{corzett} D. T. Corzett, A. M. Keller, and P. Seligmann, The thermal conductivity of manganin wire and
Stycast 2850 GT between 1 K and 4 K, Cryogenics \textbf{61}, 505, (1976).
\bibitem{peroni} I. Peroni, E. Gottardi, A. Peruzzi, G. Ponti, and G. Ventura, Thermal conductivity of manganin below 1 K, Nucl. Phys. B \textbf{78}, 573, (1999).
\bibitem{dfpt} S. Baroni, S. de Gironcoli, A. Dal Corso, and P. Giannozzi, Phonons and related crystal properties from density-functional perturbation theory, Rev. Mod. Phys. {\bf 73}, 515 (2001).
\bibitem{qe} P. Giannozzi, S. Baroni, N. Bonini, M. Calandra, R. Car, C. Cavazzoni, D. Ceresoli, G. L. Chiarotti, M. Cococcioni, I. Dabo \textit{et al.}, {\sc quantum espresso}: a modular and open-source software project for quantum simulations of materials, J. Phys.: Condens. Matter {\bf 21}, 395502 (2009).
\bibitem{pbe} J. P. Perdew, K. Burke, and M. Ernzerhof, Generalized gradient approximation made simple, Phys. Rev. Lett. {\bf 77}, 3865 (1996).
\bibitem{gbrv} K. F. Garrity, J. W. Bennett, K. M. Rabe, and D. Vanderbilt, Pseudopotentials for high-throughput DFT calculations, Comp. Mater. Sci. {\bf 81}, 446 (2014).
\bibitem{grimme} S. Grimme, Semiempirical GGA-type density functional constructed with a long-range dispersion correction, J. Comp. Chem. {\bf 27}, 1787 (2006).
\bibitem{baro09} V. Barone, M. Casarin, D. Forrer, M. Pavone,  M. Sambi, and A. Vittadin, Role and effective treatment of dispersive forces in materials: Polyethylene and graphite crystals as test cases, J. Comp. Chem. {\bf 30}, 934 (2009).
\bibitem{paukov} I. E. Paukov, P. G. Strelkov, V. V. Nogteva, and V. I. Belyi, Low-temperature heat capacity of black phosphorus, Dokl. Akad. Nauk SSSR \textbf{162}, 543 (1965).
\end{thebibliography}

\begin{thebibliography}{}
\bibitem{endo} S. Endo, Y. Akahama, S. Terada, and S. Narita, Growth of large single crystals of black phosphorus under high pressure, J. Appl. Phys. \textbf{21}, L482 (1982).
\bibitem{akiba} K. Akiba, A. Miyake, Y. Akahama, K. Matsubayashi, Y. Uwatoko, H. Arai, Y. Fuseya, and M. Tokunaga, Anomalous quantum transport properties in semimetallic black phosphorus, J. Phys. Soc. Jpn. \textbf{84}, 073708 (2015).
\bibitem{zavaritskii} N. V. Zavaritskii and A. G. Zeldovich,  Thermal conductivity of some technical materials at low temperatures, Zhur. Tekh. Fiz. \textbf{26}, 2032, (1956).
\bibitem{corzett} D. T. Corzett, A. M. Keller, and P. Seligmann, The thermal conductivity of manganin wire and
Stycast 2850 GT between 1 K and 4 K, Cryogenics \textbf{61}, 505, (1976).
\bibitem{peroni} I. Peroni, E. Gottardi, A. Peruzzi, G. Ponti, and G. Ventura, Thermal conductivity of manganin below 1 K, Nucl. Phys. B \textbf{78}, 573, (1999).
\bibitem{dfpt} S. Baroni, S. de Gironcoli, A. Dal Corso, and P. Giannozzi, Phonons and related crystal properties from density-functional perturbation theory, Rev. Mod. Phys. {\bf 73}, 515 (2001).
\bibitem{qe} P. Giannozzi, S. Baroni, N. Bonini, M. Calandra, R. Car, C. Cavazzoni, D. Ceresoli, G. L. Chiarotti, M. Cococcioni, I. Dabo \textit{et al.}, {\sc quantum espresso}: a modular and open-source software project for quantum simulations of materials, J. Phys.: Condens. Matter {\bf 21}, 395502 (2009).
\bibitem{pbe} J. P. Perdew, K. Burke, and M. Ernzerhof, Generalized gradient approximation made simple, Phys. Rev. Lett. {\bf 77}, 3865 (1996).
\bibitem{gbrv} K. F. Garrity, J. W. Bennett, K. M. Rabe, and D. Vanderbilt, Pseudopotentials for high-throughput DFT calculations, Comp. Mater. Sci. {\bf 81}, 446 (2014).
\bibitem{grimme} S. Grimme, Semiempirical GGA-type density functional constructed with a long-range dispersion correction, J. Comp. Chem. {\bf 27}, 1787 (2006).
\bibitem{baro09} V. Barone, M. Casarin, D. Forrer, M. Pavone,  M. Sambi, and A. Vittadin, Role and effective treatment of dispersive forces in materials: Polyethylene and graphite crystals as test cases, J. Comp. Chem. {\bf 30}, 934 (2009).
\bibitem{paukov} I. E. Paukov, P. G. Strelkov, V. V. Nogteva, and V. I. Belyi, Low-temperature heat capacity of black phosphorus, Dokl. Akad. Nauk SSSR \textbf{162}, 543 (1965).
\bibitem{kaneta} C. Kaneta, H. Katayama-Yoshida, and A. Morita, Lattice dynamics of black phosphorus, Solid State Commun. \textbf{44}, 613 (1982).
\bibitem{fujii} Y. Fujii, Y. Akamaha, S. Endo, S. Narita, and Y. Yamada, Inelastic neutron scattering study of acoustic phonons of black phosphorus, Solid State Commun. \textbf{44}, 579 (1982).

\end{thebibliography}
\end{document}